\documentclass[twocolumn,letter]{jpsj3}
\usepackage{txfonts,bm,color,ulem}
\title{
Entanglement Chern Number for an Extensive Partition of a Topological Ground State
}

\author{
Takahiro {\sc Fukui}$^1$ and
Yasuhiro {\sc Hatsugai}
$^2$
}
\inst{
$^1$Department of Physics, Ibaraki University, Mito 310-8512, Japan\\
$^2$Institute of Physics, University of Tsukuba, 
Tsukuba, Ibaraki 305-8571, Japan}

\recdate{ \hspace{50mm} }

\abst{
If an extensive partition in two dimensions yields a gapful entanglement spectrum
of the reduced density matrix, the Berry curvature 
based on the corresponding entanglement eigenfunction defines the Chern number.
We propose such an entanglement Chern number as a useful, natural, and calculable topological invariant,
which is potentially relevant to various 
topological ground states.
We show that it serves as an alternative topological invariant
for time-reversal invariant systems and as a new topological invariant for a weak topological
phase of a superlattice Wilson-Dirac model. 
In principle, the entanglement Chern number can also be effective 
for interacting systems such as topological insulators
in contrast to $Z_2$ invariants.
}

\begin{document}
\maketitle

The Chern number, which determines
the quantized Hall conductivity in the integer quantum Hall effect
as shown by Thouless {\it et al.} \cite{TKNN82,Kohmoto85},  
has become increasingly popular for condensed matter physicists 
owing to the recent remarkable development of the classification 
\cite{Zirnbauer96,AltZir97,SRFL08,Kitaev08}
of the topological phases of matter.
\cite{KanMel05a,KanMel05b,BerZha06,BHZ06,FuKan06,FKM07,MooBal07,QHZ08,
Roy09,FuKan07,Fu11,TIreviews}
Even in three dimensions, a type of topological insulator can be characterized by 
the Chern number, called the mirror Chern number, 
if a system has mirror symmetry.\cite{TFK08}
The numerical method of computing the Chern number has also been established \cite{FHS05} and 
widely applied to various systems with complicated multiband structures. 
The Chern number is a topological invariant for the bulk, whereas the number of edge states 
for a system with a boundary gives the same topological invariant. 
This is well known as the bulk-edge correspondence.\cite{Hatsugai93}

The entanglement spectrum of the reduced density matrix
also informs us of the edge states along an artificial boundary 
introduced by a partition of the system.
\cite{RyuHat06,LiHal08} 
Nowadays, this is widely used to clarify the property of topological insulators.
\cite{PHB10,TZV10,HPB11,AHB11,CPSV11,HuaAro12,FGB13}
Recently, Hsieh and Fu have introduced the notion of the ``bulk entanglement spectrum''. \cite{HsiFu13,HFQ14}
By considering an extensive and translationally invariant partition in real space,
they demonstrated 
that a topological ground state  intrinsically has a hidden phase transition.

In this work, we study the {\it bulk} property of some topological ground states
using the wave function of the entanglement Hamiltonian. 
Namely, we define the Chern number from the Berry curvature based on the entanglement wave function,
which is referred to as 
the {
\it
entanglement Chern number. 
}
For such a Chern number to be well-defined, 
we consider an extensive partition without a clear boundary between patches of the partition, which makes 
the entanglement spectrum gapped generically.
We then show that
this serves as an alternative topological invariant for a time-reversal invariant system.
At the same time, it yields a new topological invariant for a system with some spatial structure such as 
a superlattice system.

To begin with, let us reconsider a typical model of the topological insulator, 
the Kane-Mele model
\cite{KanMel05b}, $H_{\rm KM}$,
and exemplify the usefulness of the entanglement Chern number.
This is one of models for the quantum spin Hall effect (QSHE),
describing the electrons on the honeycomb lattice with spin-orbit couplings.
When the Rashba spin-orbit coupling vanishes, the Hamiltonian is decoupled into spin-up and spin-down sectors
such that $H_{\rm KM}={\rm diag}(H_\uparrow,H_\downarrow)$,
where $H_\sigma$ is equivalent to the spinless Haldane model for the anomalous Hall effect.
\cite{Haldane88} 
These two sectors are transformed into each other under time reversal, making
the model time-reversal invariant.
The 
half-filled ground state of this decoupled model can be characterized by two Chern numbers 
$(c_\uparrow,c_\downarrow)$, 
being a trivial insulating state when they are $(c_\uparrow,c_\downarrow)=(0,0)$,
and the QSHE state (or anomalous Hall state from the viewpoint of the Haldane model)
when $(c_\uparrow,c_\downarrow)=\pm(1,-1)$.
Although the Rashba spin-orbit coupling does not break time-reversal symmetry, it breaks the spin conservation.
Therefore, it is no longer possible to define the set of Chern numbers $c_\sigma$ for the generic Kane-Mele model.
What we can know is only the total Chern number $c=c_\uparrow+c_\downarrow$,
but it is trivial ($c=0$) because of time-reversal symmetry.

If one gives up defining the Chern number in the Brillouin zone, a spin Chern number is available
\cite{SWSH06,FukHat07a}
using a spin-dependent twisted boundary condition.
When the Rashba coupling vanishes, such a spin Chern number corresponds to $c_\sigma$, as it should be.
Even without disorder or interactions, however, 
we have to always compute it by definition in the coordinate space,
which is impossible in the Brillouin zone.

The breakthrough was the Z$_2$ topological invariant 
introduced by Fu and Kane. \cite{FuKan06}
This is, roughly speaking, half the Chern number, i.e., the integration of the Berry curvature over
half the Brillouin zone from which the Berry phase along the boundary is subtracted.
In this definition, a specific gauge fixing between the wave function of the Kramers pair
is needed. \cite{FuKan06}
The numerical method of computing the Z$_2$ invariant was also given, \cite{FukHat07b}  
which is just a straightforward generalization of the method for the Chern number. 
If the system has inversion symmetry, the above formula of the Z$_2$ invariant reduces to 
the product of the parity of the occupied bands
at the time-reversal invariant momenta.\cite{FuKan07}

In what follows, we first propose an alternative invariant for the QSHE,
the {\it entanglement spin Chern number}.
It is similar to the spin Chern number via a twisted boundary condition:
Indeed, they are manifestly equivalent when the Rashba coupling vanishes.
Moreover, the entanglement spin Chern number
can be defined in the Brillouin zone
even without the spin conservation. 
To be concrete, let us regard the spin degrees of freedom as a partition of the system. Then, tracing out 
one spin sector from the density matrix
yields an effective Hamiltonian, called the entanglement Hamiltonian, 
for the other spin sector. 
%
Since the partition in terms of the spin
is manifestly extensive and maintains 
the translational invariance, the entanglement Hamiltonian can be represented 
in the Brillouin zone, as mentioned above.  
If the spectrum of such a bulk entanglement Hamiltonian has a gap, we can define a new
Chern number different from $c(=0)$ for $H_{\rm KM}$. Here, 
note that
the entanglement Hamiltonian thus obtained  never has time-reversal symmetry. 
This implies that a nonzero Chern number can be expected in general.
In the classification of the topological phases of matter, 
symmetry protection
has surely been playing a crucial role, \cite{SRFL08,Kitaev08} but 
%
symmetry constraints are sometimes too restrictive to define the corresponding topological invariants.
Thus, the entanglement Chern number
proposed in this paper allows for various possibilities to capture
the characteristic feature of symmetry-protected topological phases.
The entanglement Chern number so far discussed is useful not only for the QSHE 
(or more generically symmetry-protected topological states)
but also some other systems
with some internal or other degrees of freedom. For example, if the system has a spatial structure such as 
a superlattice, it gives a new topological invariant, as will be discussed in the latter part of this paper.

Let $|\Psi\rangle$ be a many-body ground state of a given noninteracting Hamiltonian $H$, and let 
$A$ and $B$ be  the subsystems of the total system $A+B$.
The reduced density matrix $\rho_A$ and the corresponding entanglement Hamiltonian $H_A$ 
are defined by tracing out $B$ such that
\begin{alignat}1
\rho_A\equiv
{\rm tr}_B|\Psi\rangle\langle\Psi| =\frac{1}{Z}e^{-H_A} ,
\end{alignat}
where $Z$ is the normalization constant.
In the case of noninteracting fermions, the entanglement Hamiltonian can be written as
$H_A=\sum_{i,j\in A}c_i^\dagger {\cal H}_{A,ij}c_j$.  
Let us next introduce the correlation matrix
\begin{alignat}1
C_{ij}=\langle c_i^\dagger c_j\rangle,
\end{alignat}
where $i$ and $j$ denote the sites as well as some internal degrees of freedom such as the spin
or orbital.
When $i$ and $j$ are restricted in $A$, the correlation matrix may be called $C_A$, and 
it is shown as\cite{Peschel03}
\begin{alignat}1
{\cal H}_A^T=\ln(1-C_A)/C_A . 
\end{alignat}
Thus, the eigenstates of $C_A$ are those of ${\cal H}_A$.

As introduced by Hsieh and Fu, \cite{HsiFu13}
we consider the extensive partition with translational symmetry.
For example, in the case of the Kane-Mele model, 
we choose $A$ and $B$ as the spin-up and spin-down sectors.
Then, the Fourier transformation gives
\begin{alignat}1
C_{A}(k)=
P_AP_-(k)P_A ,
\label{EntHam}
\end{alignat}
where $P_-(k)=\psi(k)\psi^\dagger(k)$ is the projection operator to the occupied bands 
expressed by the single-particle multiplet wave functions
$\psi(k)=(\psi_1(k),\psi_2(k),\cdots )$,
and $P_A$ is the projection operator to $A$.
Solving the eigenvalue equation for $C_A(k)$,
\begin{alignat}1
C_A(k)\tilde \psi_n(k)=\xi_n(k)\tilde\psi_n(k) ,
\end{alignat}
we can define the (entanglement) Chern number as usual.
To this end, let us first consider the spectrum $\xi_n(k)$. Without the projection 
operator $P_A$ in Eq. (\ref{EntHam}), $P_-(k)$ has only two obvious eigenvalues, i.e.,  $1$ and $0$, 
denoting 
the occupied and unoccupied states, respectively. The wave functions with the eigenvalue $1$ are 
nothing but those of the ground state for the total system $A+B$.
Because of the projection operator $P_A$ in Eq. (\ref{EntHam}), 
the eigenvalues $\xi_n$ of $C_A(k)$ are not restricted to 1 and 0. Here, we assume that
some of them form bands at approximately $\xi\sim1$ and others at approximately $\xi\sim0$, and that
there is a finite gap between these two bands. 
Then, 
their origin is clear: The former are occupied states
and the latter are unoccupied states for the subsystem $A$. 
Therefore, it is natural to choose the entanglement Chern number of the {\it upper} bands
to characterize the topological property of the ground state.
In the examples we study below,  $C_A(k)$ in Eq. (\ref{EntHam})
is a $2\times 2$ matrix, and the behavior of its spectrum 
indeed exhibits such a property.\cite{footnote}

To be concrete, 
we introduce the Berry connection 
$\tilde A_\mu(k)=\tilde\psi^\dagger(k)\partial_{\mu}\tilde\psi(k)$ and the curvature 
$\tilde F_{12}(k)=\partial_1\tilde A_2(k)-\partial_2\tilde A_1(k)$,
where $\tilde\psi(k)$ is the multiplet wave function 
of the upper bands $\tilde\psi_n(k)$ with $\xi_n\lesssim1$,
and $\partial_\mu\equiv \partial/\partial k_\mu$.
The entanglement Chern number is thus defined by 
\begin{alignat}1
\tilde c_A=\frac{i}{2\pi}\int \tilde F_{12}(k)d^2k.
\end{alignat}
The above procedure is quite easy to carry out numerically using the link and plaquette
variables on the discretized Brillouin zone.
\cite{FHS05}
In this calculation of the (entanglement) Chern number, it does not depend on the gauge of 
the (entanglement) wave function even for the QSHE with time-reversal symmetry. 

Let us now compute the entanglement spin Chern number for the Kane-Mele model.
As has already been discussed, this model has two phases, the QSHE phase and trivial insulating phase.
They are distinguished by the Z$_2$ invariant or spin Chern number.  
Let us choose partitions $A$ and $B$ as the up-spin $\uparrow$ and down-spin $\downarrow$ states, respectively.
These partitions are manifestly extensive and translationally invariant.
From the numerical calculation,\cite{FHS05}
it turns out that the QSHE phase and trivial phase have
the entanglement spin Chern numbers $(\tilde c_\uparrow,\tilde c_\downarrow)=(-1,1)$ and 
$(0,0)$, respectively.
Therefore, we conclude that the entanglement spin Chern number can distinguish the phases
in time-reversal invariant classes with the conventional Chern number $c=0$.

We now discuss some details of the numerical calculations.
Although the entanglement spin Chern number indeed changes at the transition point between two phases, 
it is quite difficult to observe the gap closing in the entanglement spectrum,
at least, on the discretized lattice of the Brillouin zone that we adopted in our calculation. 
Furthermore, the entanglement spectrum $\xi_n(k)$, and thus the entanglement entropy are  
similar on both sides of the transition point, even in the vicinity of the transition point.
Such behavior of the entanglement spectrum is far from being the conventional one in a partition 
with a boundary showing the spectral flow of the edge states.
This implies that we are studying indeed
the bulk entanglement spectrum without boundaries. 
The change in the (entanglement) Chern number, however, should be due to the gap closing.
Therefore, we surmise that  
it occurs in a singular way like the delta function
at a few points on the Brillouin zone,
which is generically impossible  to observe on the meshes of a discretized Brillouin zone.
One reason why
the computation of the entanglement spin Chern number needs a rather larger number of meshes 
(order of $100\times 100$ meshes), than in the case of the Z$_2$ invariant (order of $10\times10$ meshes),   
near the transition points may be this singular behavior of the spectrum.
\begin{figure}
\begin{center}
\begin{tabular}{cc}
\includegraphics[width=0.35\linewidth]{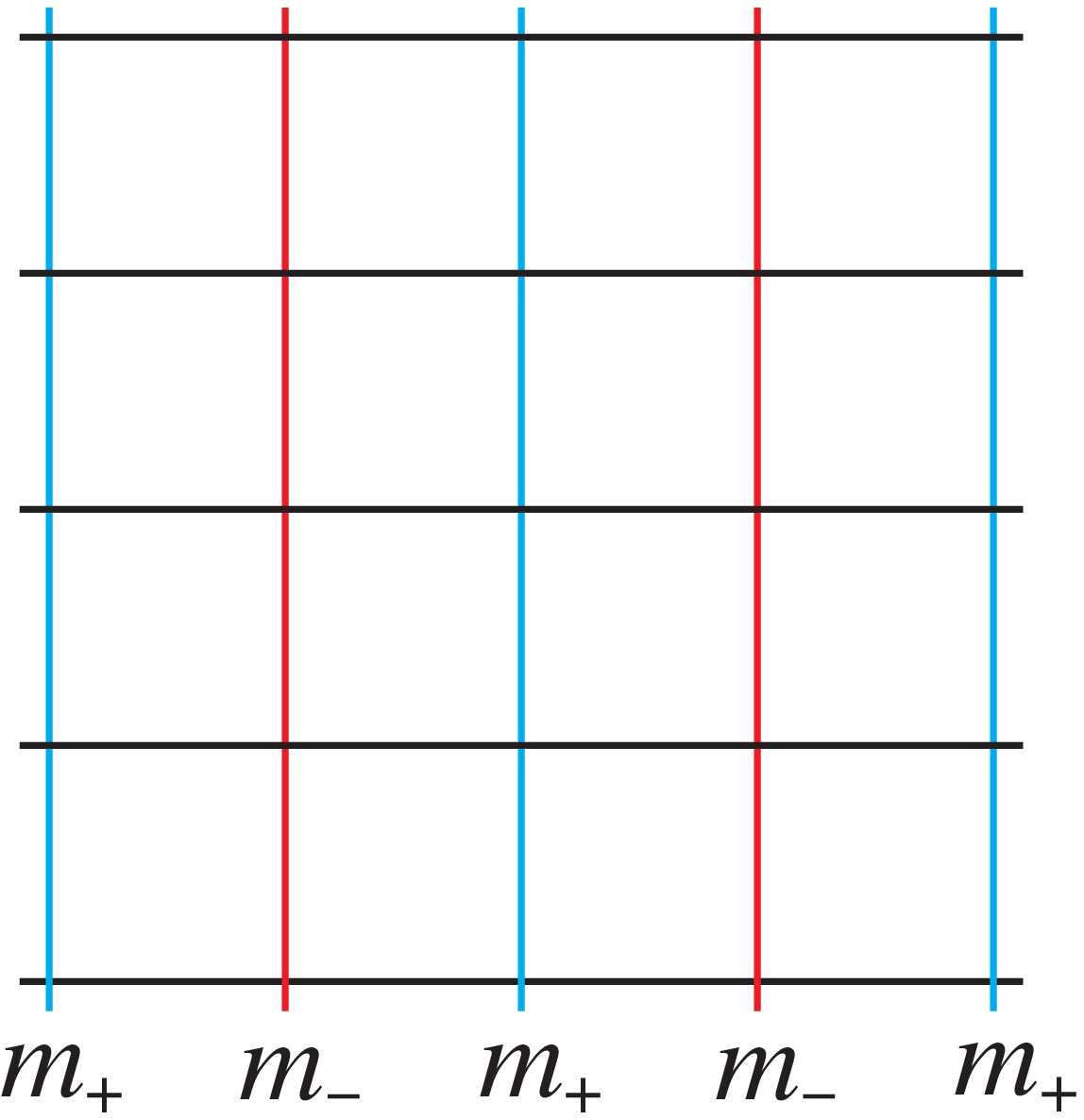}
&
\includegraphics[width=0.45\linewidth]{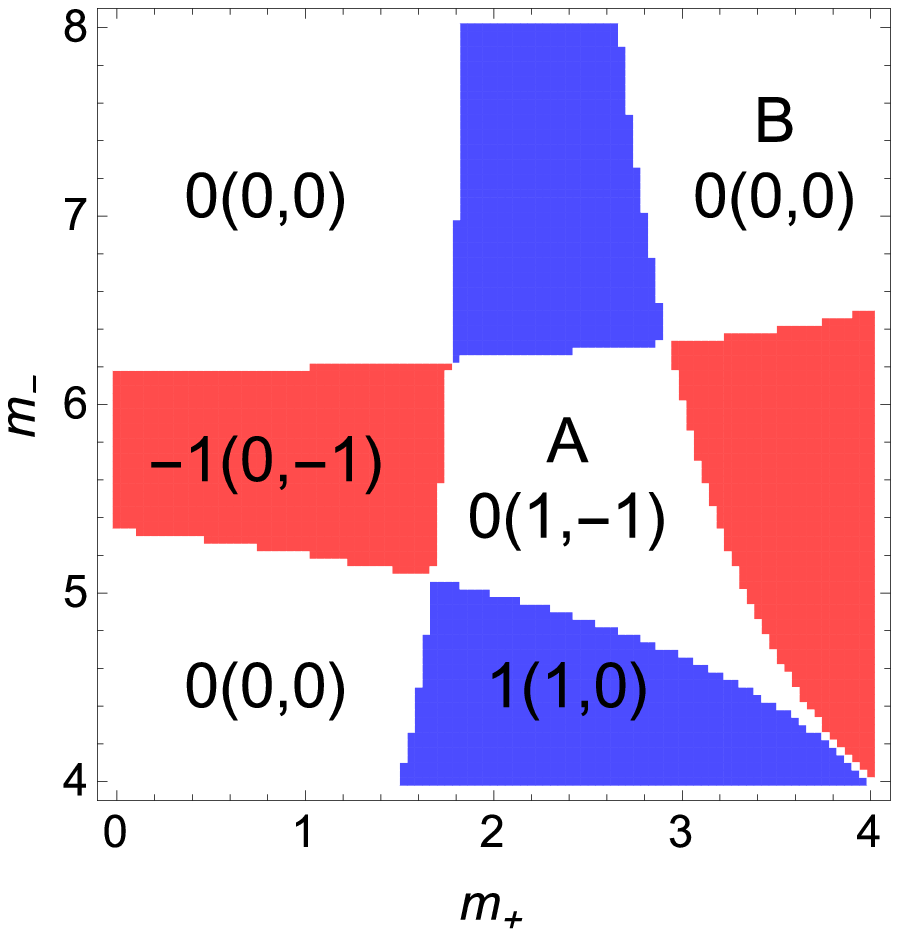}\\
(a)&(b)
\end{tabular}
\caption{
(Color online) 
(a)
Superlattice structure. On the blue and red lines, the mass parameter is set as $m_+$ and $m_-$, respectively.
(b) Phase diagram of the superlattice Wilson-Dirac model with $t=1$ and $b=2$ as a function of $m_\pm$. 
The blue, red, and white regions correspond to $c=1$, $c=-1$, and $c=0$, respectively.
A nontrivial phase with $c=0$ marked by A is given the 
entanglement Chern number dented by $c(\tilde c_+,\tilde c_-)=0(1,-1)$.
Other phases have the natural entanglement Chern numbers $0(0,0)$,
$1(1,0)$, and $-1(0,-1)$.
}
\label{f:sup_lat}
\end{center}
\end{figure}

The next example is the superlattice model of the Wilson-Dirac type. 
\cite{FIH13,YIFH14}
The Hamiltonian is given by
\begin{alignat}1
H=&\frac{-it}{2}\sum_{i,\mu} (c_i^\dagger\sigma_\mu c_{i+\hat\mu}-h.c.)+\sum_im_ic_i^\dagger\sigma_3 c_i
\nonumber\\
&+\frac{b}{2}\sum_{i,\mu}(c_i^\dagger\sigma_3c_{i+\hat\mu}+h.c.-2c_i^\dagger\sigma_3c_i),
\end{alignat}
where $\mu=1,2$ is the direction of the coordinates in two spatial dimensions, 
and $\hat\mu$ is the unit vector in the 
$\mu$ direction.
For a uniform mass $m_i=m$, the half-filled ground state has $c=1$ for $0<m<2b$, $c=-1$ for $2b<m<4b$, and $c=0$
otherwise. Therefore, we see that $m$ controls the topological phase. 
Then, what type of phase is realized when $m_i$ changes alternatively,
according to the stripe geometry shown in Fig. \ref{f:sup_lat}(a), 
where one belongs to $c=1$ and the other to $c=-1$?
As reported in Ref. \cite{FIH13}, there appears an interesting phase with nontrivial edgelike states 
despite the trivial Chern number $c=0$.
The physical situations are similar between the superlattice model and the Kane-Mele model:
A $c=0$ but a nontrivial phase is realized, 
since some degrees of freedom carry $c=1$, whereas others carry $c=-1$.  
Their coupling, however, makes it impossible to distinguish each Chern number.
The difference between the two models is in the symmetry:
In the Kane-Mele model, the nontrivial QSHE phase is protected by time-reversal symmetry.
On the other hand, there is no symmetry in the superlattice model
(or more precisely, it has particle-hole symmetry, but it does not give any protection to the phase).
This makes it difficult to characterize the nontrivial $c=0$ phase of the superlattice Wilson-Dirac model.

To overcome this difficulty, here, we propose the entanglement Chern number $\tilde c_{\pm}$
by tracing out fermions with the mass $m_{\mp}$.
We show in Fig. \ref{f:sup_lat}(b) the phase diagram as a function of $m_\pm$.
The superlattice model has the same three phases as the uniform mass model, characterized by the Chern number
$c=0,\pm1$.
However, at the center of the phase diagram marked by A in Fig. \ref{f:sup_lat}(b), 
there is a $c=0$ phase sitting just in the overlapped region 
by the $c=\pm1$ phases. As reported in Ref. \cite{FIH13}, this phase shows nontrivial edgelike states 
localized along a boundary. 
We calculate the entanglement Chern number of this phase, $c(\tilde c_+,\tilde c_-)=0(1,-1)$, which 
is manifestly distinguishable from other $0(0,0)$ states. 
In passing, we comment on another $c=0$ phase denoted by B. In this phase, 
there appear nontrivial midgap states, as reported previously.\cite{FIH13}
These states, however, can be deformed adiabatically into flatbands in the
decoupling one-dimensional 
limit $m_-\rightarrow\infty$.
Therefore, this phase has nothing to do with the entanglement, resulting in the trivial entanglement
Chern number $0(0,0)$. 

Thus far, we have shown some examples for which the entanglement Chern number is quite useful.
However, there are, of course, cases in which it plays no role.
For example, if we set $m_+=m_-=m$, the superlattice model reduces to the conventional Wilson-Dirac model
with a uniform mass term. In this case, one cannot define the entanglement Chern number by 
formally tracing out $m_+$ or $m_-$ fermions, since 
the entanglement spectrum is gapless, as proved by Hsieh and Fu. \cite{HsiFu13}
In this case, the system has an exact translational invariance by one lattice spacing, whereas, in
the case $m_+\ne m_-$, it has a translational invariance only by two lattice spacings (by one unit cell).
Therefore, the entanglement spectrum can be gapful in nonuniform mass cases.

Finally, let us mention a possible application of the 
entanglement Chern number
for interacting systems. Even though the $Z_2$ invariants of the
topological insulators are quite useful for describing
nontrivial topological phases with surface Dirac fermions, 
it cannot be applied for the interacting case, at least, directly. 
On the other hand, the entanglement Chern number can be calculated
as the ground-state Chern number of the reduced entanglement 
Hamiltonian ${ H}_A={\rm Const.}-\ln \rho_A$. By the exact diagonalization
for a finite system,
this many-body Hamiltonian can be obtained numerically 
even  with interaction.
Also, it guarantees the topological stability of the entanglement Chern number
and the corresponding topological phases against a small but finite
amount of interaction assuming that the entanglement Hamiltonian is
gapped. 
Previously, the $Z_2$ invariant for an interacting many-body system was proposed
by Lee and Ryu. \cite{LeeRyu08}
Their idea is to construct a ``Kramers doublet'' for a many-body wave function. 
To be concrete, let us consider an $N$-site system with two orbitals $t=1,2$ under the twisted
boundary conditions.
Each electron is specified by the site $r=1,\cdots,N$, the orbital $t=1,2$, and the spin 
$s=\uparrow,\downarrow$.
Let us assume a $2N$ electron ground state with an excitation gap, and consider the 
wave functions of the excited states with $N$ electrons by creating $N$ holes with $t=1$ and specific $s$ 
in the ground state. There are two independent states associated with $s=\uparrow,\downarrow$,
generically. Since the twist angle obeys the same transformation law as the momentum, 
an appropriate basis in this two-dimensional space of the many-body wave
function yields 
a ``Kramers doublet in the momentum representation''.
Here, it is crucial to assume that $N$ is odd, since the Kramers doublet does not 
occur in an even-electron system. 
The ``Kramers doublet'' thus obtained 
may be interpreted, in our description,  as the entanglement wave function by 
choosing the subsystems $A$ and $B$ as $t=1$ and $t=2$. 
Contrary to this, our proposal in this paper is 
to choose the subsystems as the spins $s=\uparrow$ and $s=\downarrow$, 
and thus we do not need to assume that $N$ is odd.
Moreover, the computational method of the Chern number is well-established 
compared with that of the Z$_2$ invariant.
These are the practical advantages of our approach.


In summary, we have introduced the entanglement Chern number defined using the 
wave function of the entanglement Hamiltonian. For the entanglement spectrum to be gapful, 
we have considered an extensive and translational invariant partition. 
We have shown that a (time-reversal) symmetry-protected topological (QSHE) phase 
can be characterized by
the entanglement spin Chern numbers $(\tilde c_\uparrow,\tilde c_\downarrow)$ obtained by 
introducing a (time-reversal) symmetry-breaking partition.
We have also shown that a $c=0$ but nontrivial phase in the superlattice generalization of the Wilson-Dirac model 
can be characterized by
the entanglement Chern numbers $(\tilde c_+,\tilde c_-)$ obtained
by introducing a natural superlattice partition.
Although the entanglement entropy and spectrum have been used to clarify
the properties of edgestates along the 
boundary between partitions $A$ and $B$, 
the entanglement Chern number proposed in this paper reveals the bulk topological properties
by an extensive partition without a boundary. 
Since the bulk-edge correspondence in real space plays an important role 
in the study of a topological state,
it is interesting to investigate
the relationship between two types of partition also 
in the entanglement description of a topological phase.


\section*{Acknowledgments}
We would like to thank K.-I. Imura and Y. Yoshimura for fruitful discussions on 
the Wilson-Dirac model.
This work was supported by Grants-in-Aid for Scientific Research (KAKENHI) Grant Numbers 
25400388 (TF), 
25610101(YH), and 26247064
form Japan Society for the Promotion of Science (JSPS).


\begin{thebibliography}{99}
\bibitem{footnote}
We point out that 
in a more complicated system that needs a more complicated 
partition, $C_A$ may become a larger matrix. In such a case, it is possible 
that the spectrum has nontrivial gaps
between multiple bands: hence,  we can basically obtain many entanglement Chern numbers.
It is then unclear which Chern number is relevant to the present purpose.
However, it is not the case in this paper, and 
we may leave the problem as a future work.
\end{thebibliography}
\end{document}